\documentclass[a4paper,11pt]{article}
\usepackage{pos}
\usepackage{subcaption}
\usepackage{hyperref}
\usepackage{placeins}      %
\usepackage[sort&compress]{natbib}
\title{Paucity of downward UHE neutrino tracks in Icecube versus unexpected huge KM3-230213A event : solving the puzzles? }
\ShortTitle{ARCA puzzling event}

\author*[, a,b]{D. Fargion}
\affiliation[a]{Rome University “La Sapienza” and MIFP, Rome, Italy.}
\affiliation[b]{Osservatorio Astronomico di Capodimonte, INAF, Naples, Italy}

\emailAdd{daniele.fargion@fondazione.uniroma1.it}

\abstract{ Recently the ARCA array detector  published  the  down-ward-horizontal event: the KM3-230213A. It appeared as the most energetic neutrino ever observed: about 200 PeV ($2 \cdot 10^{17}$ eV ) up to  EeV ($10^{18}$ eV) energy. This huge value, is puzzling. It is not statistically consistent  with several  upper bound derived by two greater and longer life  detectors: by ICECUBE and  in particular by AUGER array.  Asymmetry in recent  ICECUBE  neutrino alert tracks upward and downward at same horizontal angles  as ARCA one, suggest that they are mostly polluted  atmospheric muon bundles. This paucity also disfavor the skimming neutrino interpretation by ARCA. We suggest that the array floating and bending  in the deep sea may lead, sometime, to a misleading geometry that is pointing to a wrong arrival angle direction: a much less horizontal muon (neutrino) track respect to a much real one, more  inclined and vertical, due to atmospheric muon bundle or charmed single event. Contrary to present argument, if  such a rare  event would be  soon rediscovered in data or re-observed, it  would open the road  to  a new guaranteed Tau neutrino Astronomy.  At EeV energy such upward tau air-showers should shine  AUGER telescopes  or blaze future  satellite in Space. A  previous model in astrophysics considered   energetic $ E_{\nu} >>100$ EeV,  neutrino scattering, onto cosmic, relic, light mass ones.  Their ultra-relativistic Z boson resonance formation and its decay in flight  would produce hadron UHECR relics around  tens-hundred EeV energy. Explaining how sources located at far distances, above the usual GZK  hundred Mpc, cut off  ones, may shine and cluster in AUGER or TA data.}

\ConferenceLogo{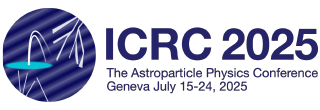}

\FullConference{39th International Cosmic Ray Conference (ICRC2025)\\
 15–24 July 2025\\
Geneva, Switzerland\\}
\graphicspath{{Figs/}{./}}
\def\orcid#1{\kern .08em\href{https://orcid.org/#1}{\includegraphics[keepaspectratio,width=0.6em]{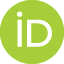}}}
\begin{document}
\maketitle

\section{Introduction} The KM3-230213A  highest energy  muon neutrino event due to its horizontal muon is puzzling. The ICECUBE cubic kilometer is a much wider  and much older array than the more recent ARCA array. It had never observed such kind of energetic tracks. Because of neutrino oscillation, a correlated upward Tau air-shower would be observed within  three thousands kilometer area  in AUGER, Argentina.  Indeed, such  muon neutrino track in KM3-230213A should have  a guaranteed companion, at hundreds of PeV, $10^{17}$ eV or even EeV $10^{18}$ energy, a tau neutrino. A necessary mixed flavor component along wide neutrino cosmic flights. Such an up-going horizontal tau neutrino had to interact sometime on the AUGER  rock surface, producing, by their up-going tau at horizons. The heaviest known  lepton \cite{perl1975evidence}. Its first interaction and its soon decay  at $D_{\tau}$ distances ($D_{\tau} = 50 m \cdot [E_{\tau}/{10^{15} eV}]$), should be observable at PeV energy in ICECUBE \cite{learned1995detecting} or first, hidden in the soil and later in the air, \cite{fargion2002discovering}, shining luminescent signals in telescopes.  At best, in AUGER, at EeV energies. But they  were never discovered by AUGER or by a comparable Telescope Array, TA. Based on the recent ARCA record, two years ago, and previous studies \cite{fargion2004tau}, one would expect nearly $30-90$ of such events in AUGER, depending on their spectra,  within   nearly 20 years of records in AUGER. Such an absence  as well as the negligible presence of  PeVs downward ICECUBE alert tracks at similar down-horizons , suggest a solution: an un-noticed, inclination of the Km3 array net under the variable deep sea currents. 

\subsection{ A brief history of the  lightest revolutionary particle, the neutrino}
Nearly 95 years ago, end 1930,  W.Pauli, contrary to N.Bohr and others,  wrote  an   amazing letter to Lise Meitner \cite{pauli1930pauli}, while being absent in a fundamental physics  meeting in Tubingen . It started with the famous words: "Dear Radioactive Ladies and Gentelman... ". Marie Curie and Albert Einstein were among the audience.  Pauli proposed  the un-expected existence of an "in-observable"  neutral particle : a  lightest evanescent  "neutron" , soon later re-named by E.Fermi, as a "neutrino".  This  idea was a necessary attempt to   overcome the apparent violation of the beta-decay energy \cite{noether1918invariante}.
Pauli was right on such particle existence, but just 26 years later,  on $1956$, the neutrino,  in disagreement with Pauli prophecy,  was  revealed. Just  a couple of  years before his death.  Later, the discovery of a new lepton, the muon, and its associated neutrinos forced B.Pontecorvo \cite{pontecorvo1947neutrino}, (and others \cite{maki1962remarks}),  to consider a neutral mixed flavor particle state. Their necessary tiny  masses  could link   different lepton components: the electron, the  muon and finally, since 1975,  thanks to M.Perl discovering,  the heaviest tau. 
The precise  neutrino mass measure suffered  a long sequence of  inexact proposals over the last decades.  Let us recall a few: on early $1980$  an apparent experimental  value of ten eV neutrino mass was widely  accepted in astrophysics and cosmology. Its  role was solving  the dark matter puzzle of the galactic halo and  the dark mass closure of the universe.   Its role was  also in agreement with the first unified particle models \cite{witten1980neutrino}. It has been  soon proposed the  possibility of such a light neutrino to reach us with a detectable time  lapse  from  an associated massless gravitational wave burst from Super-Novae, SN explosions. Rare stellar collapse event  occurring both in galactic or near LMC  or Andromeda galaxies. Explosions also occur in the tidal disturbance of the NS-NS   collapse event \cite{Fargion:1981gg},  or in a rare GW-GRB event \cite{fargion2018could}.   The neutrino particle with a mass bending at the relativistic regime does not differ much from the massless case \cite{fargion1981deflection}.  However, the  light  mass of the left-handed neutrino implied, by Lorentz transform,  the existence of a right-handed state,  with a much weaker interactions. The role of tiny Dirac-Majorana neutrino masses  in the early synthesis of nuclei \cite{fargion1984does} was noted. The    early thermal  right-handed neutrino interactions \cite{Antonelli:1981eg}, and its later multi-fluid galaxy  gravitational clustering \cite{fargion1981right}, had been  revealed. 
Such a multi-clustering  by dark matter components at different temperatures and masses \cite{Fargion:1983su} (as a right-handed  neutrino, SUSY neutralino, or other  weak  particles \cite{fargion1984mirror}), their different  gravitational  clustering times and masses,  might be, incidentally,  related to the very recent (unexplained)  earliest galaxy populations \cite{mann2023james}, and the early massive Black Hole presence \cite{jacak2025possible}. \cite{fargion1984may}.
Since 1980  several authors have published  unexpected claims on the neutrino mass: for instance, on 1985 J.J Simpson  claimed the presence of a heavy,  17 $keV $ neutrino mass component \cite{simpson1985evidence}.  This   possibility was in severe tension with the astrophysical and cosmological  data. Soon  it  was  dismissed.  Later on, following  the dozen of SN 1987 neutrino cluster  events, most models were  favoring  much lighter neutrino masses. The Kamiokande  atmospheric neutrino  mixing tested on 1998  and the SNO experiment on  solar neutrino flavor  oscillations and mixing both  decreased the hopes for such a dominant neutrino cosmic role. They suggested a very light $ <1$ eV neutrino mass. But it opened the road to the $\tau$ neutrino Astronomy \cite{fargion2002discovering}.  The eventual role of such a cosmic  relic eV  neutrino mass, in a hot dark halo, as an effective  beam dump for ZeV ($10^{21}$ eV)  cosmic UHE neutrinos had been advocated \cite{fargion1999ultra} to solve the  new  not explained UHECR , Ultra High Energy Cosmic Ray, $E> EeV$,  event noted on 1995.  Apparently they  correlated with very far AGN hadron sources, at distances  above the so called GZK cut off \cite{1966JETPL...4...78Z}\cite{greisen1966end}.  Most recent signatures of UHECR lightest nuclei composition,  their additional "Hot Spot" clustering, point today mainly to local AGN. They represent and fit better a nearby local origin  of UHECR \cite{fargion2024uhecr}. A more recent and  precise cosmology  models needed lightest $<1 $ eV, neutrino masses.  Above this value , around half of the Z boson mass, the eventual fourth neutrino role has been considered by several authors  in the past decades, all within astrophysical and cosmic constraints.  A fourth lepton family model is still  waiting  for  experimental confirmations \cite{belotsky2008may},\cite{belotsky2003invisible}.   In conclusion  late discovery \cite{perl1975evidence} of  the  heavier, $\tau$ , completed the standard  elementary particle frame:  six lepton (three charged, three neutral) and  six quark. 
The three neutrino  production  in terrestrial atmosphere are mainly made by muon and electron ones;  however, their flavor mixing at large galactic and cosmic distance flight guaranties the comparable presence for the rarest tau neutrino, as abundant as other ones. Such an astrophysical $\nu_{\tau}$  and its eventual consequent tau, which being so much unstable, behave, at GeV-TeV as a unique cascade event. However, at energy above hundred TeV or PeV and above , the $ \nu{\tau}$ interaction in ice and its tau decay are well separated, offering, in principle,  a characteristic double bang  in underground signature in ICECUBE \cite{learned1995detecting} \cite{fargion2015not},\cite{lad2025neutrino}. 
In addition, any energetic tau ($E_{\nu_{\tau}} >>PeV$) escaping from a mountain or rising from Earth's soil  can decay in the air, producing an upward-flowing tau air-shower \cite{fargion2002discovering}, (often incorrectly referred to as the  skimming neutrinos \cite{feng2002observability}) . Such hundreds of PeV  $ 10^{17}  eV- 10^{18}  eV$  energetic tau neutrinos could be well be observable in AUGER or in Telescope Array detectors \cite{fargion2004tau}. As rare up-going air-shower.
As we shall see in the present article, their absence at EeV energies implies a serious bound to the ARCA discovery.
 \subsection{CERN-OPERA 2011 events: the faster than  light neutrino?}
A more  surprising historical claim on the neutrino nature occurred in September 2011:  OPERA  and CERN experiments  declared that they observed a  neutrino signal flight that was  faster than light. An imaginary neutrino mass acting as a tachion.  Two days  \cite{fargion2012inconsistency} and two weeks \cite{cohen2011pair}  later,  the claim had been confuted by theoretical arguments.  The Supernovae 1987A  neutrino burst timing and the solar neutrino mixing, among electron and muon flavors, did not allow  such a result. Only six months later, more accurate tests from ICARUS \cite{antonello2012search} confirmed the respect of special relativity.  The  OPERA group, found  that a  tiny time delay  led  to the wrong experimental interpretation. Therefore, the special relativity  still applies to neutrinos, as it should. Any  small geometrical error may disturb and fake even most great experiments.
\subsection{ The ARCA 2023 KM3-230213A : the highest energetic neutrino ?}
The past decades of ICECUBE and  of AUGER   array records  are  statistically in tension with the unexpected huge 2023 neutrino event KM3-230213A in ARCA  \cite{km3net2025observation}.  The contradictions are numerous \cite{li2025clash}.
Tau air-showers  by EeV neutrinos  in AUGER should be well  observable:   last two decades  records in AUGER assuming the  nominal KM3-230213A event rate,  would produce about $70$ events \cite{fargion2004tau}. No one had been discovered yet.  The huge  ICECUBE  mass detector  and its life time  exposure  overcome the ARCA one at least  by an order of  magnitude.   This sounds statistically un-probable.
An EeV  neutrino has an ideal origin: to be a  relic \cite{beresinsky1969cosmic} of the opacity \cite{greisen1966end} ,  \cite{1966JETPL...4...78Z}, due to the decay of the UHE  Delta barion resonance into pions. This candidature  is   not  much  consistent with such KM3-230213A event.  Indeed , assuming a common cosmic UHECR  spectra,  the most recent  of the AUGER,  composition  data, \cite{mayotte2025measurement} had shown that the proton presence above a few  $ 10^{18}  eV$ , energy is negligible.
Consequently, the Delta resonance in GZK cutoff is no longer tuned with a proton energy at $6 \cdot 10^{19}  eV $ energy and with the present cosmic  $T_o =2.7  K^o$ thermal bath.  One should  require a hotter cosmic background to be  hit by a few $ 10^{18}  eV$ proton. This may take place only at a minimal temperature  , at least as $ T_{z=10} > 27  K^o$, much hotter than the present.  This  may  require early cosmological stages,  at a redshift $ z> 10$.  However, the  same later cosmic expansion  would once again reduce the final  energy of the  GZK  neutrino secondary, reaching us, in  our days,  with an energy $ E{\nu} < 10^{16}  eV$   and not  at the (observed) and expected  $ E{\nu} > 10^{17}  eV$  ones.  There are nevertheless independent processes capable of accelerating the AGN jet and, within the same photons and proton beam flow, producing UHECR  Delta resonances. This processes has been recently suggested , at smaller space and energy scale, for the micro-quasar precessing jet, such as the galactic SS433 one \cite{fargionss433}.  

\section{ The paucity of downward versus upward  ICECUBE  neutrino alert events}
\begin{figure}[th]
\centering
\includegraphics[width=0.6\textwidth]{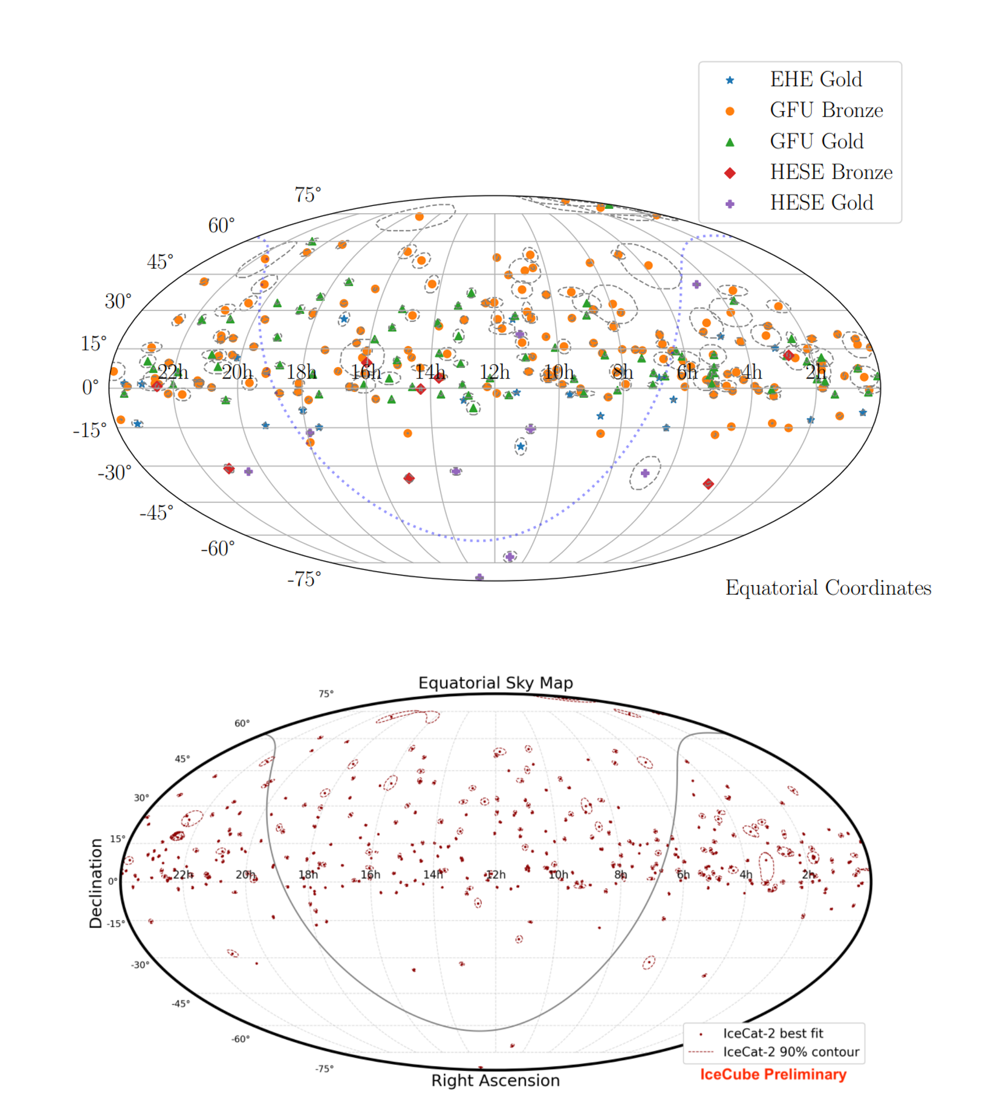} 
\caption{The ICECUBE alert tracks recorded in last years in celestial coordinate- The up-side of each figure point to the North of the Sky. Consequently  most event in South Pole array are coming from the North ( up-going) , while much  less are reaching from the South (down-going tracks). The figure above,  based on ICE-Cat1  catalog, refer to the late 2024 data \cite{abbasi2024icecat}. The figure below is based on the up dated recent but preliminary ICE-Cat2 catalog of events, shown in 2025 \cite{zegarelli2025icecat}
 The two figures differ by a minor angular definition and by  25 additional events in the 2025 ICECat2 ; \cite{zegarelli2025icecat}}
\label{fig:1}
\end{figure}

\begin{figure}[th]
\centering
\includegraphics[width=0.6\textwidth]{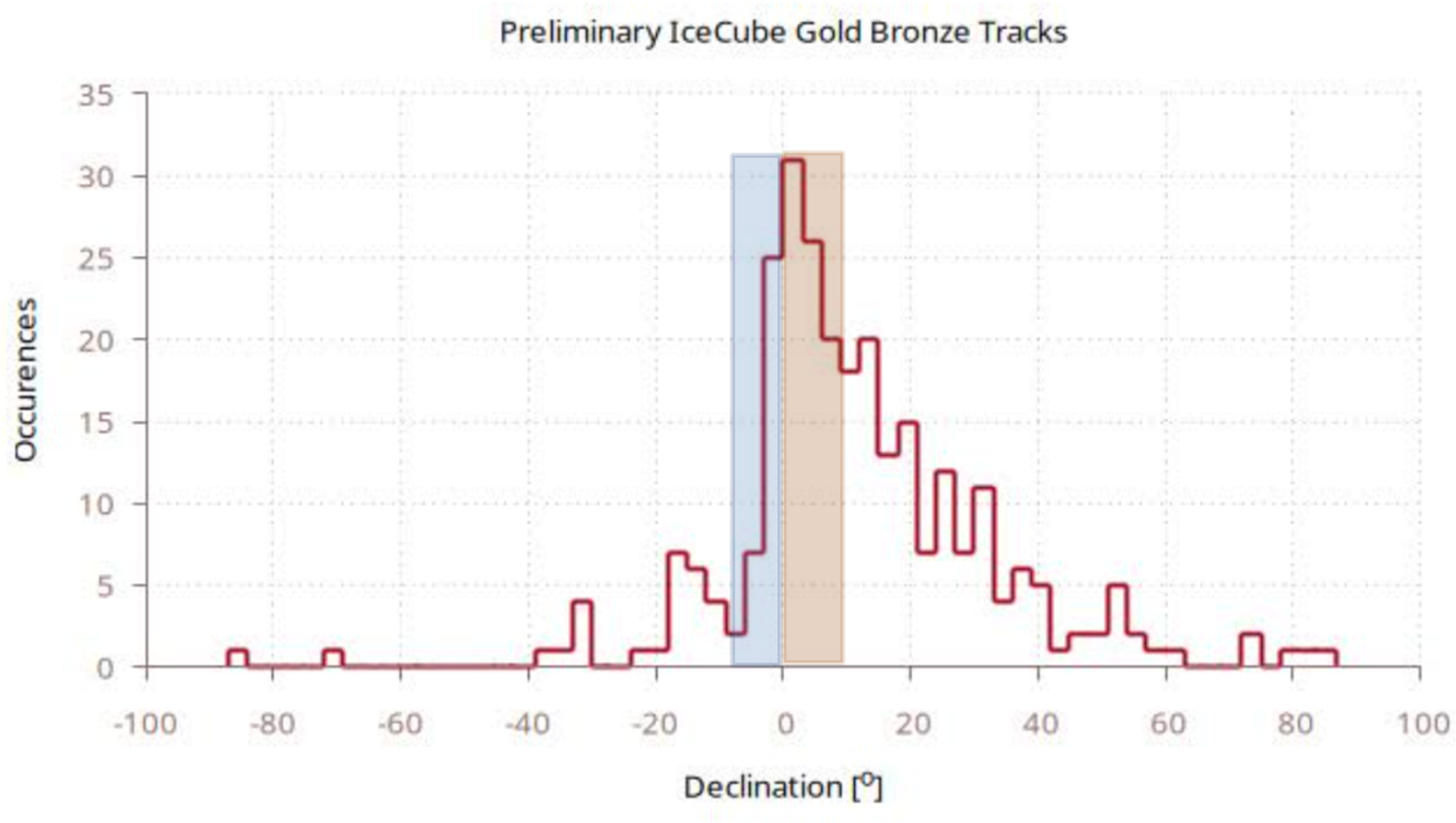} 
\caption{ The asymmetry among the $+/- 9^o$ event tracks. The paucity of downward tracks imply that the  atmospheric noise is  greatly polluting most of these horizontal-downward events, at similar arrival angle as the ARCA Km3 event, KM3-230213A.}
\label{fig:2}
\end{figure}

Among nearly 274  events in 2023 ICECUBE alert tracks at maximal energy, most alerts are  upward ones, see  Fig  \ref{fig:1}.   This asymmetry must  be related to a severe veto for such downward muon tracks, mostly of atmospheric nature. Such a veto has to reduce the over-abundant downward cosmic ray noise. This filter is also obtained  by the on surface ICETOP array. This surface array is not present in the $km^3$ sea array.  At horizons (+/-9°) where the downward and upward beam dump masses are very comparable,  see  Fig.\ref{fig:2},   the asymmetry is within 111 events , only 34  point down , while 77 are upward:  the probability P to occur by chance is quite rare: only  $P = 1.5\cdot 10^{-5} $. 

This suggests that such horizontal,  down-ward inclined tracks in ICECUBE  are mostly undesired atmospheric muon tracks. This must also be true for the Mediterranean sea $km^3$ array. The  more extensive sky asymmetry for all alarm events ,  $ 274$, is mainly $213$ , pointing upward events while only $61$   are downward ones.  For such a rate in a binomial case, the probability of happening by chance is as low as   $ P= 3.3  \cdot 10^{-9}$.See Fig. \ref{fig:2} 

\begin{figure}[th]
\centering
\includegraphics[width=0.6\textwidth]{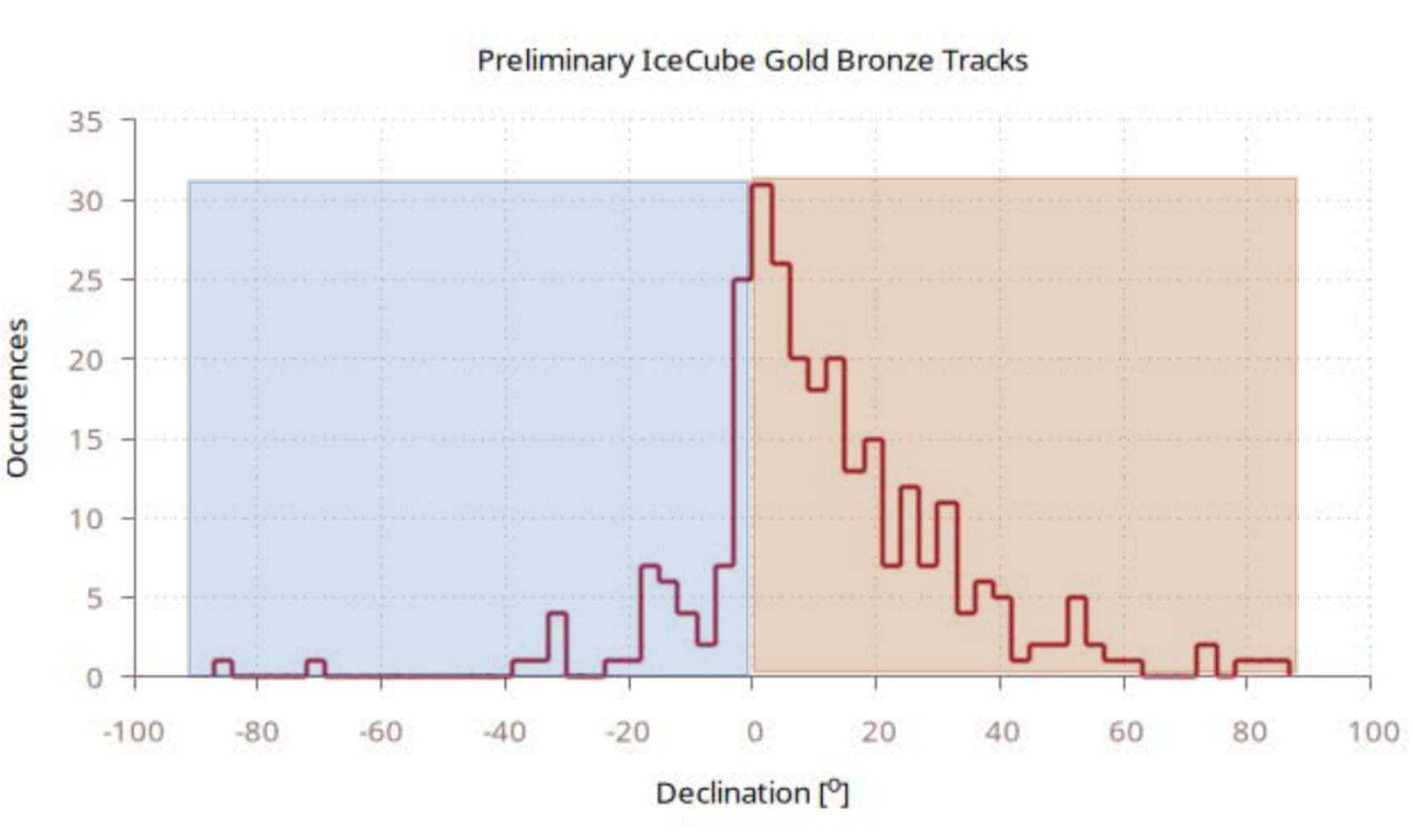} 
\caption{The asymmetry among the up and down ward event tracks. The paucity of downward tracks,( binimial probability $P$ to occur by chance  is about $ P= 3.3 \cdot 10^{-9}$, imply  that the atmospheric noise and its filtering is excluding most down-ward tracks as polluted ones.}
\label{fig:3}
\end{figure}
Recall that a muon neutrino at an energy above $30$ TeV energy should suffer absorption  while crossing the whole Earth. Therefore, this asymmetry in the entire sky , see Fig.\ref{fig:3} is  more meaningful than that shown in Fig.\ref{fig:1}.  For  the whole up/down sky alert ICECUBE event, until 2024, observed  by  ICECUBE,  274 tracks, most of them 213 , are pointing up-ward. Only 61 down-ward. The binomial probability $P$ that occurs by chance  is about $ P= 3.3 \cdot 10^{-9}$.   The Earth opacity for such hundred TeV neutrino  up-going tracks implies an additional  upward neutrino   asymmetry. 
All of  these critical considerations suggest an eventual misleading interpretation of the data. But  they do not  offer  any solution to the puzzle.

\subsection{ A key difference:  a frozen  static ice array versus a floating in sea one ?}
The main difference between ICECUBE and the ARCA sea array is the medium in which the optical elements are positioned.  ICECUBE holds the array elements within a  frozen, static, ice in the South Pole.
The ARCA  $km^3$ array is located in deep sea, where the detector elements are anchored in $3$ km depth in the soil.  The array is floating , while connected in columns, bounded deep in the water.   The sea water is not static as ICECUBE ice. Therefore, variable inclination is the difference.
Nevertheless, the  KM3-230213A event was, in general, real,  bright, and sharp. 
The normalized  number of photons in KM3-230213A was  $N_{PMT} = 3,672$ . 
Any deep sea current may blow into the  array, it may bend, and curve their element position. changing the array profile.  Imagine, for example,  the Pisa tower inclination.  A few degree bending can offer a distort view of the array coordinate system.
The KM3-230213A.  was  apparently only 0.8° degree down-ward:  just where ICECUBE excluded most of its  events, by  its filter veto for atmospheric noise ; see Fig. \ref{fig:2}, \ref{fig:3}.   Among nearly 274  events in 2023 ICECUBE alert tracks at maximal energy,  most of them  are  upward.  There must  be a severe veto, due to atmospheric muons, for such downward tracks.  This veto  was also possible thanks to the IceTop  array filter.   At horizons (+/-9°) where the downward and upward beam dump masses are comparable, the asymmetry is , within 111 events 34 down , 77 upward: the probability of such an occurrence,  by chance, is quite rare: $1.5 \cdot 10^{-5}$. Therefore, the rise of an UHE neutrino in such a horizontal downward sky is quite puzzling.
\begin{figure}[th]
\centering
\includegraphics[width=0.7\textwidth]{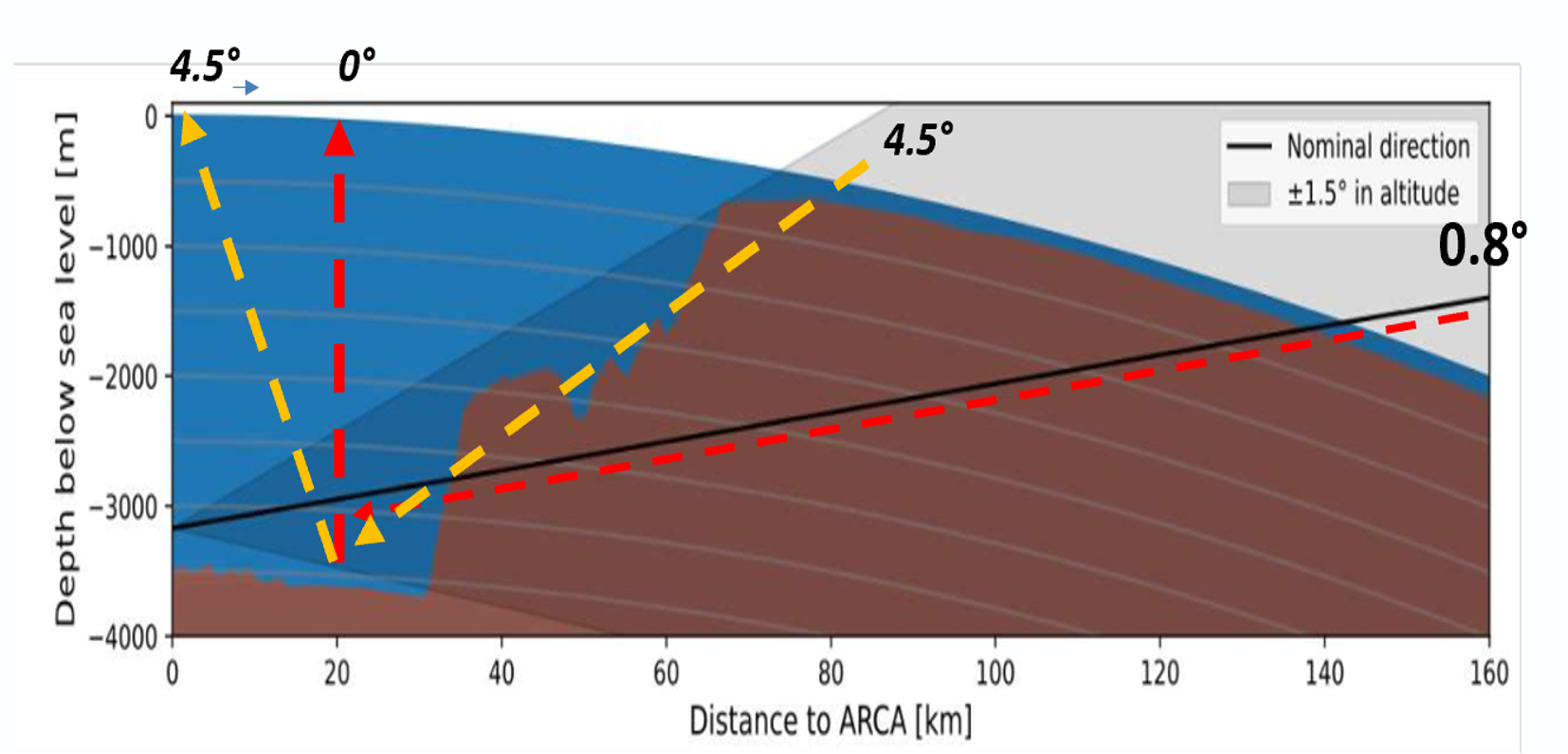} 
\caption{An exaggerated geometry showing a  downward-inclined charmed   muon  track (yellow large dashed line) . The alternative  neutrino track  (red smaller dashed line) following ARCA  event interpretation \cite{km3net2025observation}. The horizontal version  (standing for a neutrino) versus  a more inclined, vertical geometry.   This  inclined  geometry could offer an atmospheric muon interpretation}
\label{fig:4}
\end{figure}

\begin{figure}[th]
\centering
\includegraphics[width=0.7\textwidth]{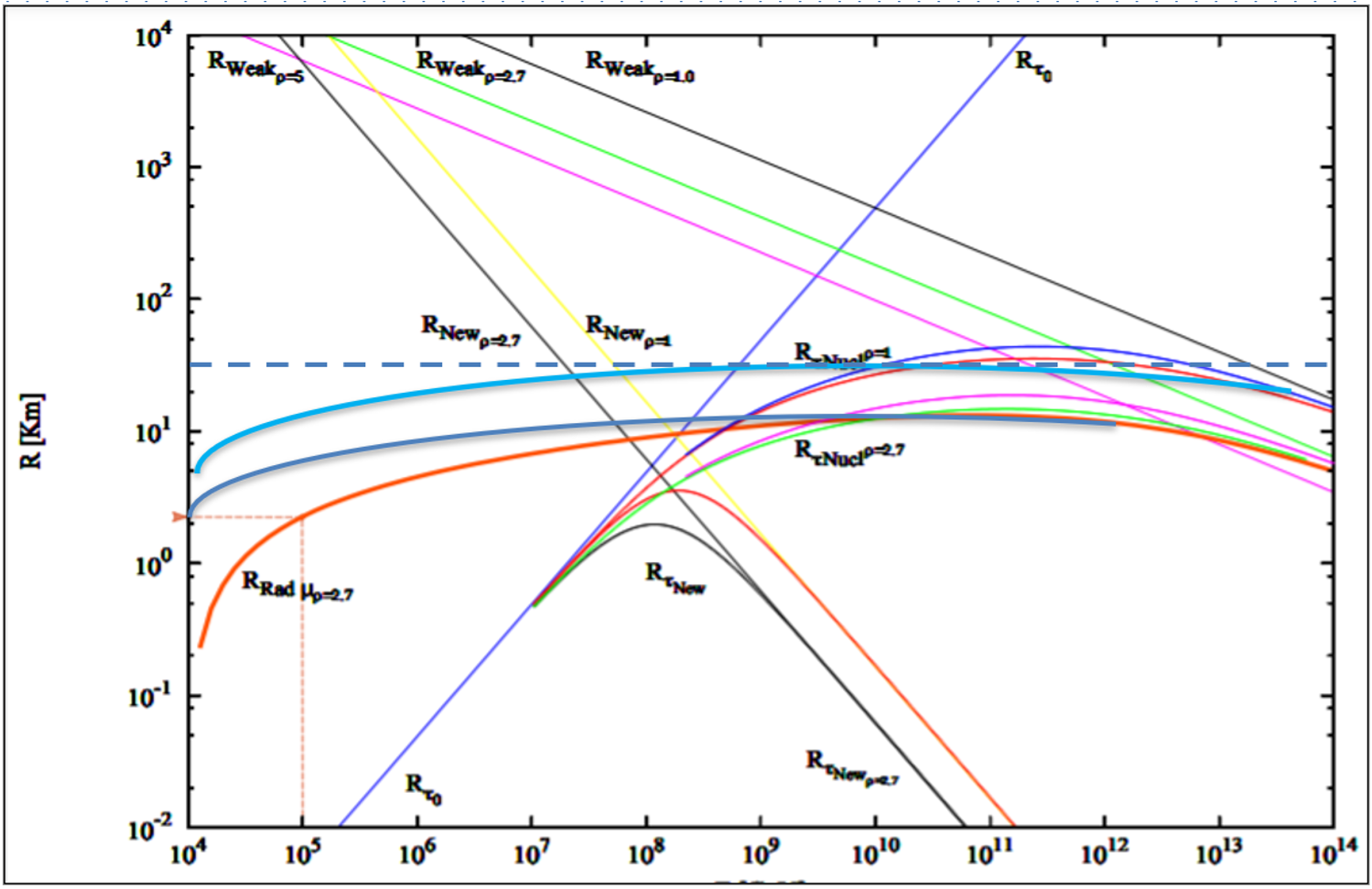} 
\caption{The different expected survival distances for muon and tau  assuming their corresponding energies. The  orange range curve has been here updated by a light blue curve,  based on  earlier articles:\cite{fargion2002discovering},\cite{fargion2004tau}.   At $\theta > = 7^o$, or at least  $\theta > 5^o$ inclination, an  energetic , EeV, atmospheric charmed  muon might reach,  the deep  sea,  respectively, nearly $20$ up to $28$ km distances, within the muon survival track distance. Easier possibility   to  reach at high energy  for  a more  bent array  $\theta > > 7^o$ }
\label{fig:5}
\end{figure}

\begin{figure}[th]
\centering
\includegraphics[width=0.7\textwidth]{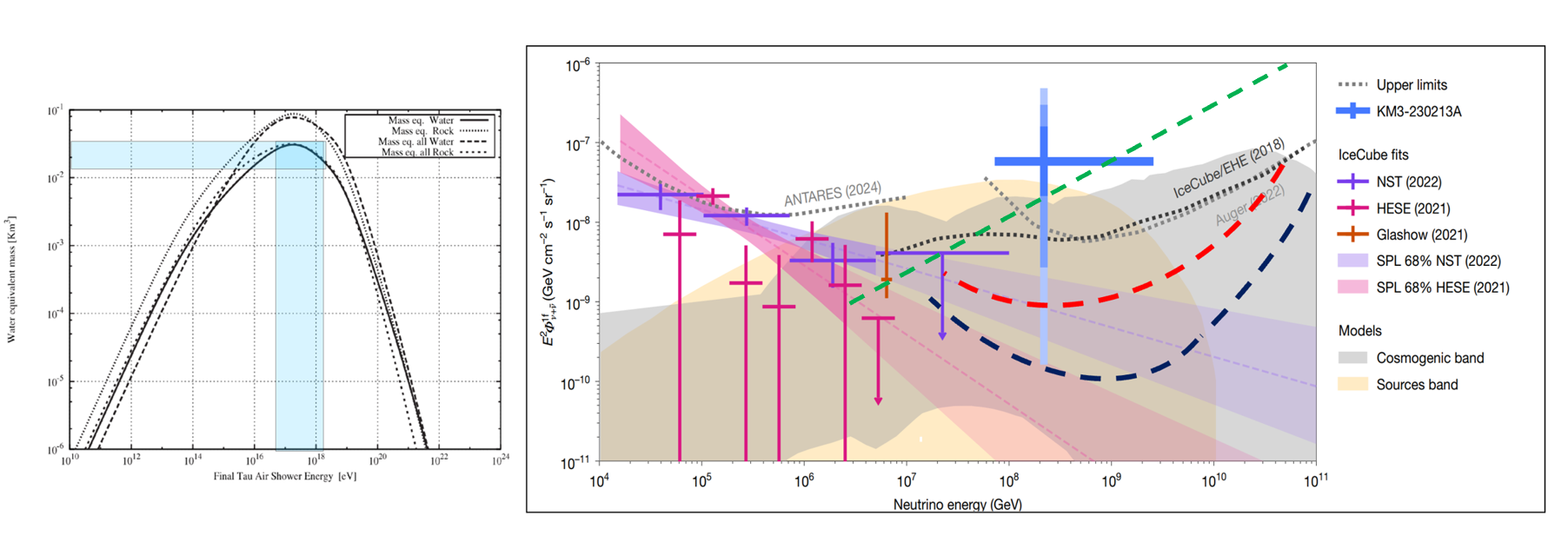} 
\caption{ On the left the  detector mass, in $km^3$ water equivalent,  for each $km^2$ area in AUGER  \cite{fargion2004tau}. On the right, the averaged EeV neutrino event by ARCA ,within previous spectra data and bounds, following \cite{km3net2025observation}. We underline, in partial disagreement with  their figure above,  that the AUGER bound in shouldbe  more effective and  more severe than their gray dot curve. :At EeV energy, at least three or four times than ICECUBE one, even considering the night $10\%$  reduction for telescopes.  Therefore the  curve on the left is transferred as a more restrictive bound (dashed red line) on the right. The united ICECUBE and the AUGER limit is shown by a lower dashed black curve.  One note that the ARCA spectra tendency  in present figure shows a trend  (dashed green line)  to  even a larger flux at ZeV energy.  Ideal and tuned for Z- Boson model \cite{Fargion:1999ss}.  Recent UHECR could be  quite successful correlated with a dozen of  Local Sheet AGN sources \cite{fargion2024uhecr}.  However  rare UHECR event clustering,  with far AGN as 3C 454 may need such ZeV neutrino courier. \cite{fargion2024uhecr}.}
\label{fig:7}
\end{figure}

\section{A pragmatic Conclusion : the critical view  }
The most probable  solution of the ARCA puzzle is that the rare bending  (in deep sea by sudden current) of the ARCA vertical array into a more inclined one led to a misunderstanding. A downward muon bundle event, several dozens of PeV muons at angles ( $\theta >> 5^o-7^o$), reaching  above the horizons, might be relics of a hadron  UHECR event originated on inclined-horizons, hundreds kilometer away. 

\subsection{A speculative Conclusion : an exciting options }
If the ARCA  EeV neutrino event is more verified and   confirmed as a real one, it might  skipped to  wider ICECUBE detector discovery because of their severe filtering of very horizontal arrival directions, to avoid atmospheric noises. The AUGER array detector , three-four times more  capable in the EeV energy windows,  could not observe such an up-going air-shower if their real extreme energy is above tens of EeV energy.   Indeed, such extreme neutrino energy  $ E_{\tau} > 10^{19} eV$, will produce a decay distance too far $\tau$:     ($D_{\tau} > 500 km \cdot {(E_{\tau}/(10 EeV)))}$. :   the only surviving events will be very horizontal. Their decay will be extremely far , at high, rarefied  altitudes . These levels of diluted atmosphere  are so low that  the  consequent airshower slant depth is not large enough to form  bright fluorescence light , easily observed. This  energy will be an opportunity for different tuned theoretical model to revive and confirm. The tuned  energetic  neutrino events at highest ,  ZeV, $10^{21}$ eV energy , are  a  necessary  and required  tool  in the Z-Burst model.  Model suitable for solving some rare  UHECR puzzles.
Better studies on ICECUBE horizontal muons should  reveal such rare signals.  Anyway, AUGER array might better review their up-going air-shower signatures.
Such a Z-Burst model is capable of explaining far (above GZK) UHECR arrival  sources  by their scattering onto our  dark matter galactic relic neutrino halo, with $0.1-0.4 $ eV  mass, in  extended  Mpc halos. 
Time and wider detectors would confirm,  with any additional events,  also based also on tau air-showers, the nature of such  exciting  EeV neutrino astronomy. 
No new signal would slowly favor the inclined array, which misled the interpretation of the ARCA event.
More  EeV neutrino signals,  their mass splitting combined with UHECR spectra  bumps and clustering, might offer a road  map  to revolutionary neutrino mass spectroscopy \cite{fargion2001shadows} .

\subsection{ Dedication}
The present article is dedicated to the   tau discovery by M.Perl half century ago,  to the energy conservation theorem by Emmy Noether more than a century ago,  to the nuclear fission understanding by Lisa Meitner,  to the proposal  of a neutrino mixing by  Bruno Pontecorvo,
and  in a Jewish humorist  letter, the sudden vision for an  invisible neutrino,  by  Wolfgang Pauli .

\bibliographystyle{JHEP}
\bibliography{daf2025v25}

\end{document}